\documentclass[]{spie}  

\usepackage[pdftex]{graphicx}
\DeclareGraphicsExtensions{.pdf,.jpeg,.png}

\usepackage{amsmath}
\usepackage{amsfonts}

  \usepackage[caption=false,font=footnotesize]{subfig}

\usepackage{url}

\usepackage{cite}

\newcommand{\z}{\mathbf{z}}
\newcommand{\x}{\mathbf{x}}
\newcommand{\w}{\mathbf{w}}
\renewcommand{\t}{\mathbf{t}}
\newcommand{\boldmu}{\mbox{\boldmath{$\mu$}}}

\newcommand{\R}{\mathbb{R}}

\def\ig{\includegraphics}

\newcommand{\Pbkg}{P_{\mbox{\scriptsize bkg}}}
\newcommand{\Ptgt}{P_{\mbox{\scriptsize tgt}}}

\newcommand{\D}{\mathcal{D}}
\renewcommand{\L}{\mathcal{L}}

\newcommand{\ie}{\textit{i.e.}}

\newcommand{\eq}[1]{Eq.~(\ref{eq:#1})}
\newcommand{\fig}[1]{Fig.~\ref{fig:#1}}

\usepackage{xcolor}
\newcommand{\remark}[1]{\textcolor{blue}{[\textit{#1}]}}
\renewcommand{\remark}[1]{}

\title{Sculpting priors}

\author{James~Theiler
  \skiplinehalf
  Space Remote Sensing and Data Science Group,\\
  Los Alamos National Laboratory,\\ Los Alamos, NM 87545, USA
}
\authorinfo{Email: jt@lanl.gov}

\begin{document} 
\maketitle 

\begin{abstract}

Bayesian priors are investigated for detecting targets of known spectral signature (but unknown strength) in cluttered backgrounds.  A specific problem is the construction (or ``sculpting'') of a Bayesian prior that uniformly outperforms its non-Bayesian counterpart, the nominally sub-optimal but widely used Generalized Likelihood Ratio Test (GLRT).
  
\end{abstract}


\remark{
}


\section{Introduction}
\label{sect:intro}

This study is motivated by the problem of detecting targets
in cluttered backgrounds.  The specific example considered here is the
detection of solid sub-pixel targets in multispectral (or
hyperspectral) imagery\cite{Schaum97,Manolakis05,Matteoli14,Manolakis16,Theiler18igarss,Ziemann18,Besson20,Vincent20a,Theiler20arxiv}.
Although there are many variants and real-world complications,
a simple statement of the problem considers a target with spectral
signature $\t$ and abundance (\ie, size relative to a pixel) $a$, against
a background $\z$, leading to a measured pixel value
\begin{equation}
  \x = (1-a)\z + a\t
  \label{eq:replacement-model}
\end{equation}
Here, $\t$, $\z$, and $\x$ are $d$-dimensional vectors (elements of $\R^d$),
corresponding to the $d$ spectral channels of the imaging sensor; and
$a$ is a scalar bounded between 0 and 1.

In this expression, $\t$ is known (it is typically measured in the
laboratory, or directly from the sensor at locations where the target
is known to cover a full pixel), and although the background $\z$ is
not known in detail, it is assumed to be drawn from a distribution
$\Pbkg(\z)$ that is known (it is usually inferred from measurements over
a large number of presumably [or at least mostly] target-free
pixels).  In terms of the target abundance $a$, we can write the
distribution associated with the measured pixel:
\begin{equation}
  \Ptgt(a,\x) = (1-a)^{-d}\Pbkg\left(\frac{\x-a\t}{1-a}\right).
\end{equation}
Note that $\Ptgt(0,\x) = \Pbkg(\x)$.

The basic goal, here, is to find targets -- in the case of imagery, to
identify the pixels in which the targets appear.  Specifically, we
seek a function $\D(\x)$ that maps the observed vector $\x$ to a
scalar that correlates with target presence.  Targets are identified
by comparing $\D(\x)$ to a threshold $\eta$.  Values of $\D(\x)$ above
the threshold correspond to targets, and values below the threshold
are taken to be target-free.  A bit of terminology: the real-valued
function $\D(\x)$ is called a ``detection statistic'' while the binary
function given by comparing $\D(\x)$ to a threshold~$\eta$ is called a
``decision rule.'' The term ``detector'' is used informally (and sometimes
ambiguously) to refer to one or the other (or sometimes both).

There are (at least) two different approaches for obtaining a
detection statistic $\D(\x)$: estimation and hypothesis testing.  An
\emph{estimator} is a function $\widehat{a}(\x)$ that aims to
approximate the true abundance~$a$.  A typical (but not the only
available) choice is the maximum likelihood estimator
\begin{equation}
  \widehat{a}(\x) = \mbox{argmax}_a \Ptgt(a,\x).
\end{equation}

If we want to know if and/or where targets might be in a given image
(or corpus of images) we could do worse than computing $\widehat
a(\x)$ over all the pixels, and then taking as targets those pixels
for which the value is above some threshold. But we could also do
better.

The hypothesis testing approach aims to more directly infer whether
$a$ is zero or non-zero.  For the \emph{simple} hypothesis testing
problem, there are only two alternatives: $a=0$ and $a=a_0$ for some
$a_o>0$ that is specified beforehand.  This is sometimes referred to
the ``clairvoyant'' case\cite{Kay98}. It is a conceptually useful
case, even as it is unrealistic in those cases for which the nonzero target
strength~$a_o$ is \emph{not} known \textit{a priori}. What is useful about it
is that it has an unambiguously
optimal solution, given by the likelihood ratio:
\begin{equation}
  \D_c(a_o,\x) = \frac{\Ptgt(a_o,\x)}{\Pbkg(\x)}.
  \label{eq:likelihood-ratio}
\end{equation}

\subsection{Measures of detector quality}

A good detector will find most of the targets with only a few false alarms.
Thus we seek to maximize the detection rate (DR)
\begin{equation}
  \mbox{DR} = \int_{\D(\x)\ge\eta}\Ptgt(a,\x)~d\x
\end{equation}
while minimizing the false alarm rate (FAR)
\begin{equation}
  \mbox{FAR} = \int_{\D(\x)\ge\eta}\Pbkg(\x)~d\x.
\end{equation}
Note that both of these quantities depend on the threshold~$\eta$. Decreasing $\eta$ will generally increase the detection rate, but at the expense of also increasing the false alarm rate.  A plot of DR vs FAR as $\eta$ varies defines the so-called Receiver Operating Characteristic (ROC) curve.  The \emph{power} of a detector is its detection rate at the threshold $\eta$ that corresponds to a given false alarm rate.  The area-under-the-curve (AUC) statistic corresponds to the area under the ROC curve -- it is a sort of average power, with the average taken over all false alarm rates from 0 to 1.  Another criterion that is often of interest in detection problems is the false alarm rate at a fixed detection rate.

It is important to observe that DR depends on the target strength~$a$.
A consequence of this is that it is possible for one detector to be
better (that is, to be more powerful) than another detector at one value
of~$a$, and worse than that other detector at some other value of~$a$.
A \emph{uniformly} more powerful detector will be better at all values
of $a$.  (To be a little more precise: a uniformly more powerful
detector will be as good or better at all values of~$a$, and strictly
better at at least one value of~$a$.)

For a known fixed target strength $a=a_o$, the optimal detector
is given by the likelihood ratio in \eq{likelihood-ratio}.

In practice, however, $a$ may not be a known fixed quantity.  In fact,
there are two senses in which $a$ is not known: 1/ it is not known
whether the target is present in the given pixel, so we do not know if
$a$ is zero or nonzero; and 2/ presuming $a$ is nonzero, we do not
know which nonzero value it has.  Because our ignorance of $a$ extends
into this second case, our problem becomes one of \emph{composite
hypothesis testing}, and we cannot use use the simple likelihood ratio
in \eq{likelihood-ratio}.

The non-Bayesian approach is to employ the likelihood ratio using the
maximum-likelihood estimator for $a_o = \widehat{a}(\x)$.  This leads to
the Generalized Likelihood Ratio Test (GLRT):
\begin{equation}
  \D(\x) = \frac{\mbox{max}_a \Ptgt(a,\x)}{\Pbkg(\x)}.
\end{equation}
The Bayesian approach is to identify a prior $q(a)$ and to average the
likelihood with respect to that prior:
\begin{equation}
  \D(\x) = \frac{\int \Ptgt(a,\x)q(a)~da}{\Pbkg(\x)}.
  \label{eq:bayes-detector}
\end{equation}

A classic result in the theory of hypothesis testing\cite{Lehmann05}
is that Bayesian detectors (strictly speaking: decision rules)
are \emph{admissible}, and that all
admissible detectors are either Bayesian or can be expressed as a limit of
Bayesian detectors. A detector is admissible if no other detector is
uniformly more powerful. It follows that for any given detector
$\D(\x)$ there must exist a prior function $q(a)$ such that the
Bayesian detector associated with that prior (that is, the Bayesian
detector defined in \eq{bayes-detector}) is uniformly as powerful or
more powerful than $\D(\x)$. This motivates our interest in constructing
a prior that enables the Bayesian detector to equal or beat any non-Bayesian detector.

As a side note, there is a further class of detectors\cite{Chen98,Schaum10,Theiler12spie,Schaum16b} that generalize the
the GLRT by incorporating a function $q(a)$ that is superficially similar
to the prior in \eq{bayes-detector}.  Here,
\begin{equation}
  \D(\x) = \frac{\mbox{max}_a \Ptgt(a,\x)q(a)}{\Pbkg(\x)}.
\end{equation}
These are not Bayesian detectors (because the ``max'' is not the same as an integral), and therefore may not be admissible, but the $q(a)$ does provide a
flexibility that is not available in the standard GLRT. \remark{so what?}

\subsection{Interpreting the prior: detectors vs estimators}

\begin{flushright}
  {~}\\
  ``It is a capital mistake to theorize before one has data.
  Insensibly, one \\ begins to
  twist facts to suit theories, instead of theories to suit facts.''\\
  -- Arthur Conan Doyle, {\sl A Scandal in Bohemia}.
  \end{flushright}

A popular interpretation of the Bayesian prior is that it  represents
an initial data-free guess (some would say: belief) for what values a
parameter might take on.  There is an inherent subjectiveness in this
interpretation (perhaps the very kind of subjectiveness that Sherlock
Holmes warns against), and a premium is placed upon priors that are
broad, flat, and uninformative. In this view, the worst mistake of all
is to employ a prior that is a delta-function about a single value.

But it is important to recognize that this interpretation of the prior
is only appropriate for \emph{estimation} problems. For example, if
our aim is to estimate $a$, and our prior is $\delta(a-a_o)$, then the
posterior estimate will always be $\widehat a = a_o$, regardless of
what data are observed. (Put another way: no matter what we observe,
our conclusion will coincide with our initial prejudice.)  On the face
of it, it is not obvious that detection and estimation are all that
different. Indeed, it is not a terrible idea (it is usually
sub-optimal, but not terrible) \remark{as noted earlier: one could do
  worse...} to detect targets by estimating a candidate target's
strength, and then checking to see if that estimate is sufficiently
nonzero.

But in fact, estimation is different from 
hypothesis testing.  And for detection via hypothesis testing,
the prior need not be
a subjective guess, but instead can express objective criteria.  Given
a specific criterion to be optimized, one can turn
around\footnote{Indeed, the very term ``turn around'' implies
(correctly) that this is a kind of inverse problem.} and construct (or
``sculpt'') the prior that optimizes this criterion.  It also bears
remarking that for the detection problem, a delta-function prior is
not necessarily a bad idea (indeed, it is the basis of the
\emph{veritas} algorithm\cite{Theiler21veritas}).

\section{Sculpting priors}

The task we have set out for ourselves is to identify a prior function
$q(a)$ with the property that it permits the associated Bayesian
detector to uniformly outperform the GLRT detector.  We have seen
previously\cite{Theiler23igarss} that this is not always possible for
some performance criteria (eg, FAR at fixed DR), but for other
criteria (eg, DR at fixed FAR; or AUC), it is guaranteed to be
possible by the theory\cite{Lehmann05}.

In the full problem, $a$ is allowed to vary continuously from 0 to
1. To enable numerical exploration of optimal priors, we will restrict
the prior to be a ``delta comb.''  That is, for a fixed number $K$,
the prior is given by $K$ weights $\w=[w_1,\ldots,w_K]$ and
$K$ values
of $a$ (called ``knots''): $\mathbf{a} = [a_1,\ldots,a_K]$. 
\begin{equation}
  q(a) = \sum_{k=1}^K w_k\delta(a-a_k).
  \label{eq:delta-comb}
\end{equation}
The use of delta-comb priors is a restriction, not an
approximation.  The Bayesian detectors that employ these priors are
computed exactly (to within numerical precision), and the resulting
decision rules are, in a strict and formal sense, admissible.

The \emph{veritas} algorithm\cite{Theiler21veritas} is a special case
in which $K=1$ and the value of $a_1$ is carefully chosen. Here, we will
consider larger values of $K$ and let $a_k$ uniformly fill the
interval; in particular, we'll divide the unit interval into $K$ equal
segments and place the $k$th knot at the midpoint\footnote{We avoid
$a=0$ since that corresponds to no target, but in fact, we could
consider a term corresponding to the $a\to 0$ limit, using an
approach described in Ref.~[\citenum{Theiler22homeopathic}].} of the
$k$th segment: thus, $a_k = (k-1/2)/K$.

In the evaluation of these delta-comb priors, we will consider only
the values of $a$ that are part of the delta-comb itself.  That is,
the problem itself has been modified so that the unknown target
abundance~$a$ is assumed to be either zero (no target) or one of the
delta-comb values in $\mathbf{a}$.  Not only does this simplify the
numerics, but it means that in the context of the problem, we are
considering \emph{all} possible Bayesian priors.  Given this
restriction, we will also consider a restricted version of the GLRT
(that we'll call the RGLRT) that also only considers the abundances in
$\mathbf{a}$.  To be more specific:
\begin{align}
  \D_{\mbox{GLRT}}(\x) &= \max_{a\in[0,1]} \frac{\Ptgt(a,\x)}{\Pbkg(\x)} \label{eq:glrt},\\
    \D_{\mbox{RGLRT}}(\x) &= \max_{a\in\mathbf{a}} \frac{\Ptgt(a,\x)}{\Pbkg(\x)} \label{eq:rglrt}.
\end{align}
For small values of $K$, we find that the RGLRT outperforms the
GLRT. For larger values of $K$, the two detectors become more nearly
equal and exhibit more nearly equal performance.  A practical
distinction between the two is that GLRT is typically evaluated by
taking the derivative of the likelihood ratio and setting that
derivative to zero, thus leading to a single closed-form expression.
The RGLRT solution is also, technically speaking, a closed-form
expression, but its evaluation requires the computation of $K$
separate likelihood ratios; thus, especially for larger $K$, the RGLRT
can be more computationally expensive.

Since we are specifically looking to find Bayesian weights that outperform the associated
(R)GLRT detector, we will seek to optimize the loss function
\begin{align}
  \L(\w) &= \max_a L(a,\w), \mbox{where}\nonumber \\
  L(a,\w) &= s_{\mbox{\scriptsize Bayes}}(\w,a) - s_{\mbox{\scriptsize RGLRT}}(a)
  \label{eq:loss}
\end{align}
where $s$ is a ROC-based performance statistic (with smaller $s$ being
better), such as FAR@DR=0.5, or 1-DR@FAR=0.05 or 1-AUC. If we can find
$\w$ such that $\L(\w)\le 0$, then that implies that $L(a,\w)\le 0$
for every~$a$, which implies that the Bayesian detector has uniformly
matched or outperformed the GLRT-based detector. 

\section{Experiments}

We perform numerical experiments with a large sample of points
drawn from simulated elliptically-contoured multivariate
$t$-distributed data. This distribution can be expressed in closed-form
\begin{equation}
  \Pbkg(\x) = c\left[\nu-2 + (\x-\boldmu)^TR^{-1}(\x-\boldmu)\right]^{-(\nu+d)/2}
  \label{t-distn}
\end{equation}
where $c$ is a normalizing constant, and $\nu$ is the ``degrees
of freedom'' parameter -- small $\nu$ corresponds to a fat tail, and
$\nu\to\infty$ leads to a Gaussian distribution. Under this
distribution, the GLRT can also be expressed in closed
form\cite{Schaum97,Theiler18igarss}, as can the clairvoyant
detector\cite{Theiler21veritas}, and therefore any Bayesian detector
with a finite delta-comb prior.

We use zero mean ($\boldmu=0$) and unit covariance ($R=I$).
(This does not really cause any loss of generality, since our
detection algorithm would be whitening the data anyway -- we'll be
using very large sample size, so the covariance would be accurately
approximated.) We take $\nu=3$, which indicates a
very non-Gaussian fat-tailed distribution.

In the first batch of experiments, $d=9$ spectral channels, $N=10^7$
target-free pixels, and $N=10^7$ corresponding pixels for each nonzero target
strength~$a$ in the delta-comb.  Note that these are matched-pair
pixels\cite{Theiler13mpml}, with each target-present pixel matched to
a specific target-absent pixel in which target has been added.
As a way of assessing variability in the estimate of the prior weight
vector $\w$ (which will
depend on the particular sample of $N$ points), we run 45 trials, and
each trial is plotted separately in the figures.

\begin{figure}
  \ig[width=0.33\textwidth]{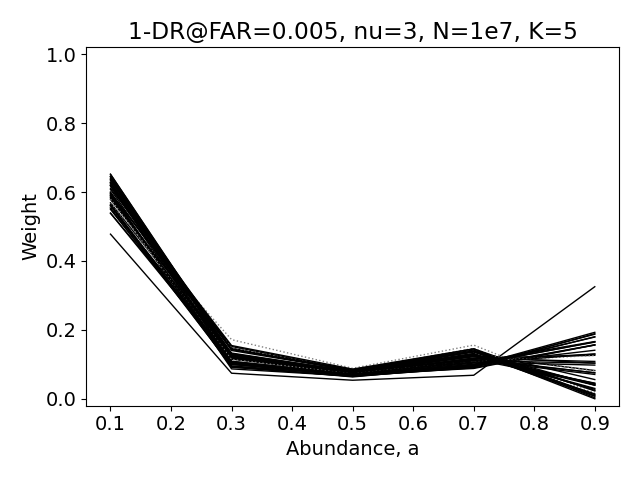}
  \ig[width=0.33\textwidth]{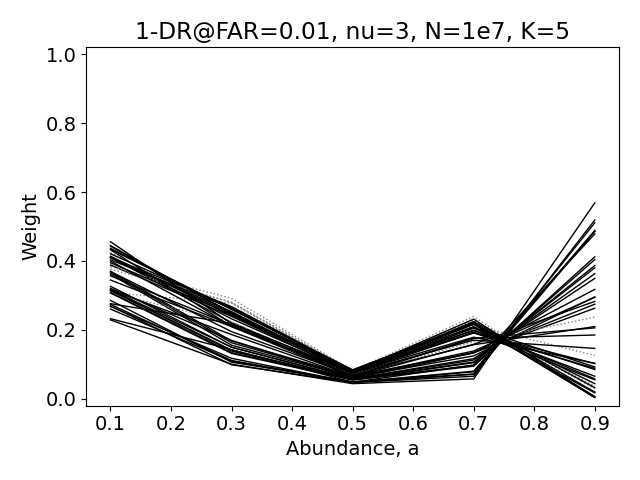}
  \ig[width=0.33\textwidth]{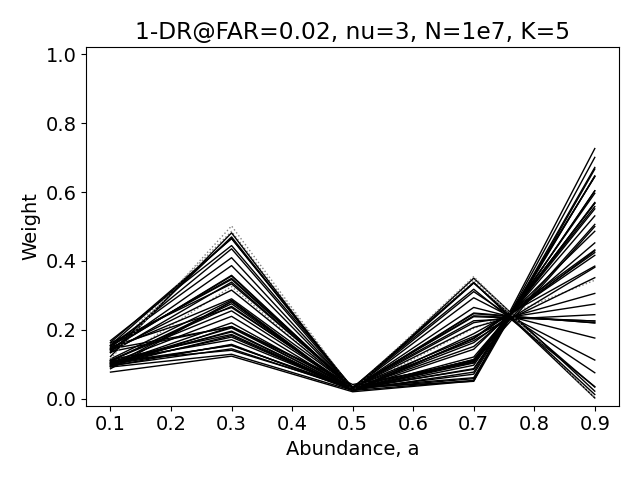}\\
  \ig[width=0.33\textwidth]{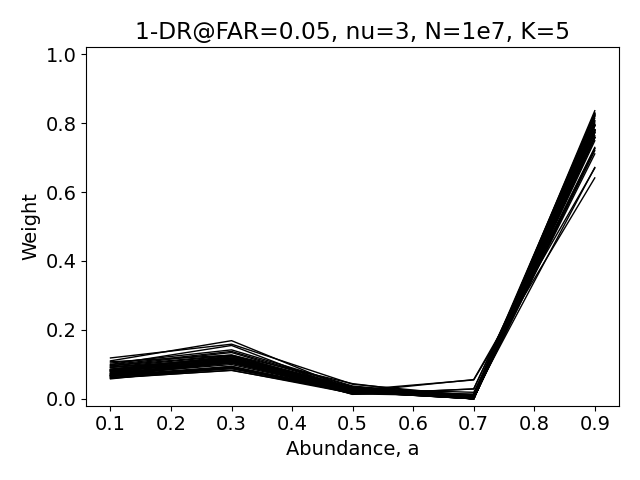}
  \ig[width=0.33\textwidth]{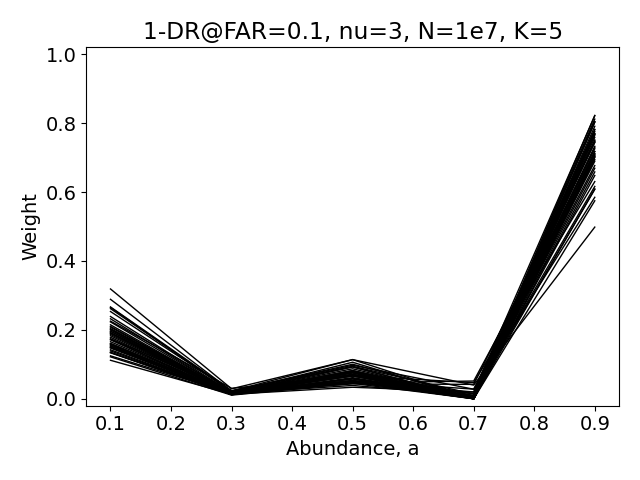}
  \ig[width=0.33\textwidth]{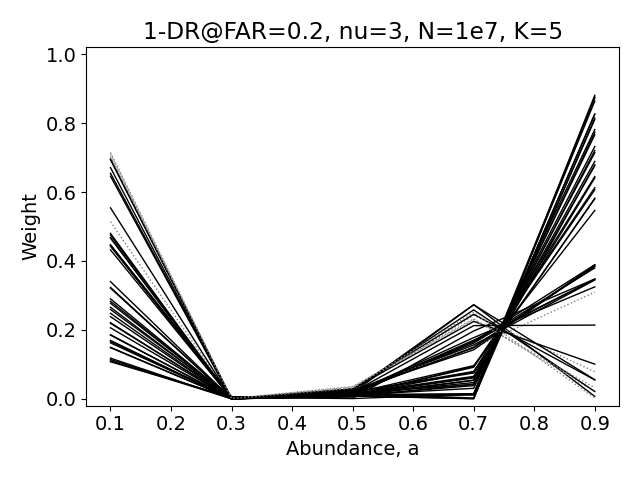}\\
  \ig[width=0.33\textwidth]{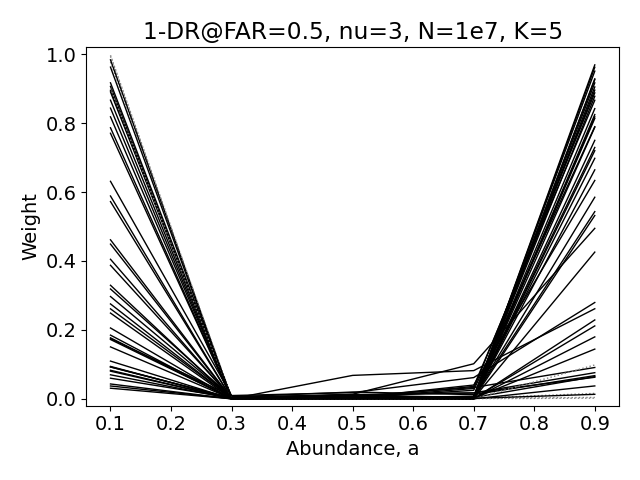}
  \ig[width=0.33\textwidth]{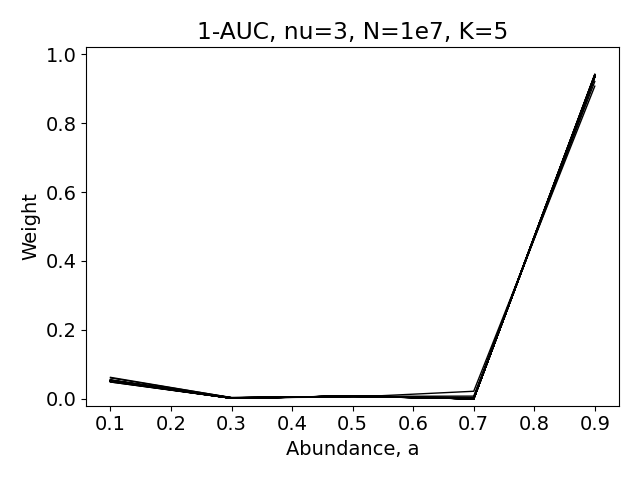}
  \ig[width=0.33\textwidth]{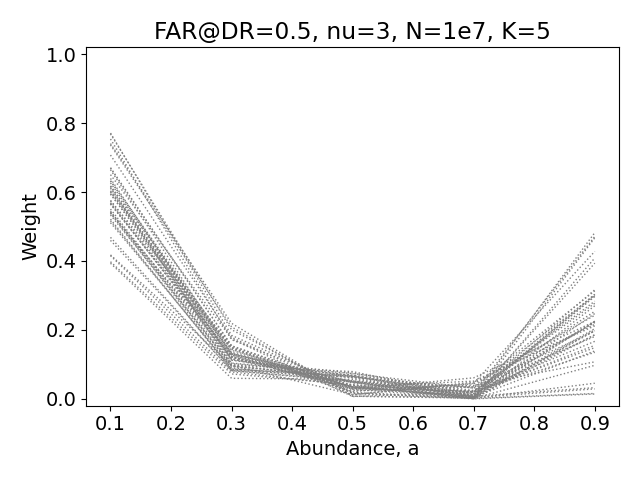}
  \caption{Plot of weights $[w_1,\ldots,w_K]$ against abundance values
    $[a_1,\ldots,a_K]$, with $K=5$, for the delta-comb prior defined
    in \eq{delta-comb}.  These weights optimize the loss function in
    \eq{loss}, as computed with $N=10^7$ samples, and based on
    different ROC-based statistics.  The first seven panels are of for
    DR@FAR=$x$ for various (increasing) values of $x$. The last two
    panels show results for the AUC statistic and for
    FAR@DR=0.5. Results are shown from 45 trials, with solid lines
    indicating solutions with $\L(\w)\le 0$ and dotted lines for
    solutions with $\L(\w)>0$.  Note that almost all of
    the lines are solid in these plots, \emph{except} for the
    FAR@DR=0.5 statistic in the last panel.  This is consistent with
    the theory that guarantees the existence of priors for which
    Bayesian detection is uniformly at least as powerful as any given
    detector, when evaluated with AUC or the DR-based statistics.  The
    theory does not apply to the FAR-based
    statistics\cite{Theiler23igarss}; in this case there appears not
    to be any prior that leads to uniform superiority, but other cases
    have been found (see \fig{priors-FAR-D144}) in which the Bayesian detector is uniformly better. \label{fig:priors-K5}}
\end{figure}

\begin{figure}
  \ig[width=0.5\textwidth]{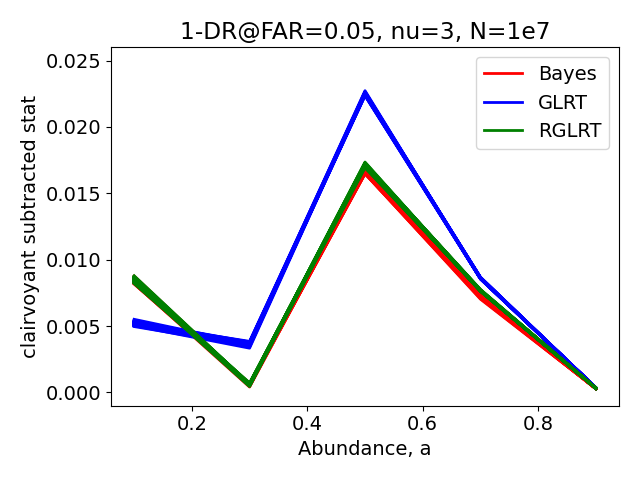}
  \ig[width=0.5\textwidth]{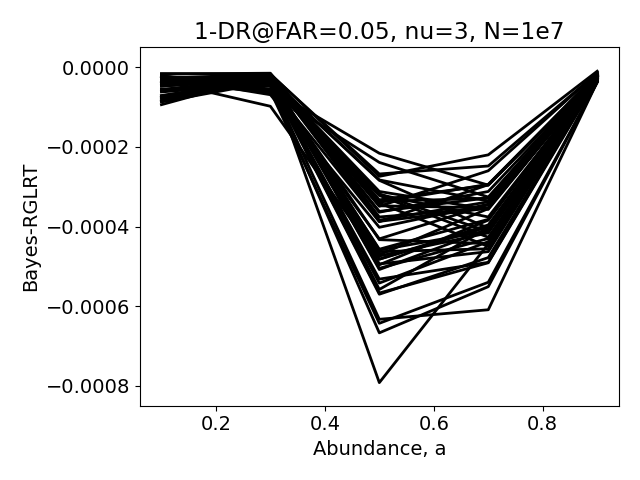}
  \caption{Performance differences: Left panel shows 1-DR@FAR=0.05
    statistic for Bayes, GLRT, and RGLRT statistics, with the
    clairvoyant performance subtracted.  We observe that these
    differences are all positive, which demonstrates the known result
    that the clairvoyant detector is optimal (though in practice it is
    usually unavailable). Right panel shows the difference of Bayesian
    and RGLRT detectors.  Here all values are negative, which
    indicates that the Bayesian detector is uniformly more powerful
    than RGLRT.  It bears remarking that the magnitude of these
    differences is small. These methods are all within a few percent
    of the clairvoyant detection rate, and the difference between Bayes
    and RGLRT is less than a tenth of a percent, often much
    less.\label{fig:differences}}
\end{figure}

\begin{figure}
  \ig[width=0.33\textwidth]{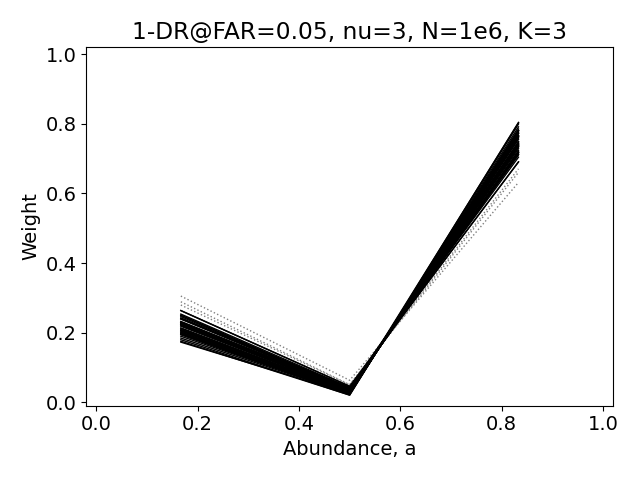}
  \ig[width=0.33\textwidth]{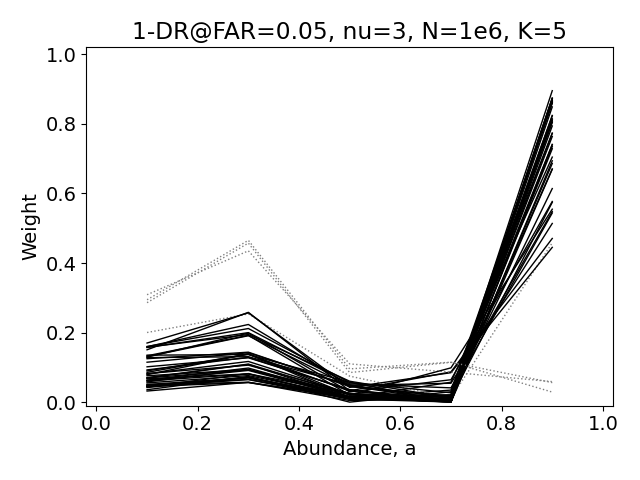}
  \ig[width=0.33\textwidth]{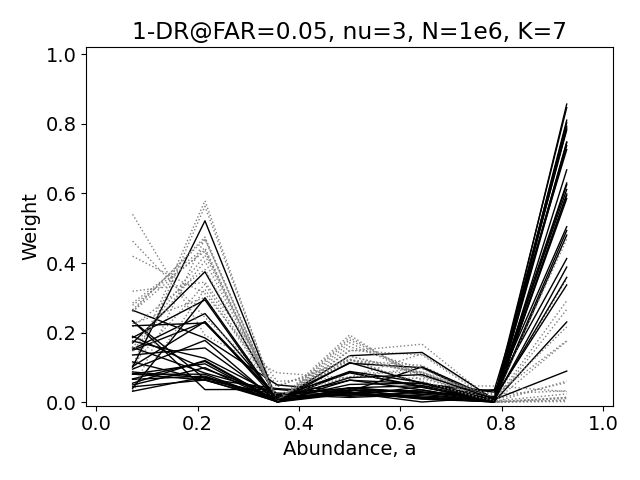} \\
  \ig[width=0.33\textwidth]{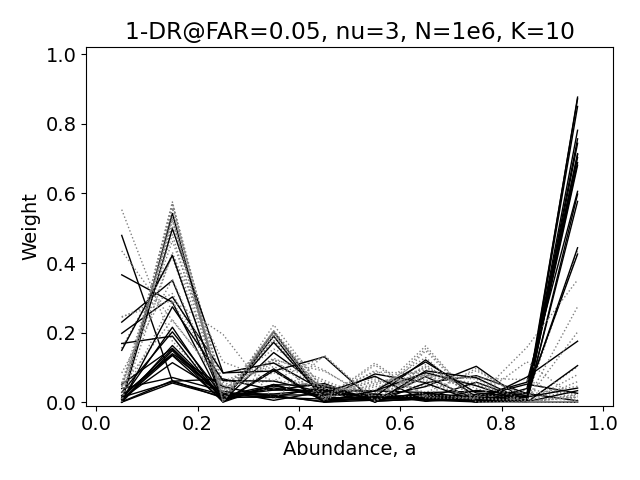} 
  \ig[width=0.33\textwidth]{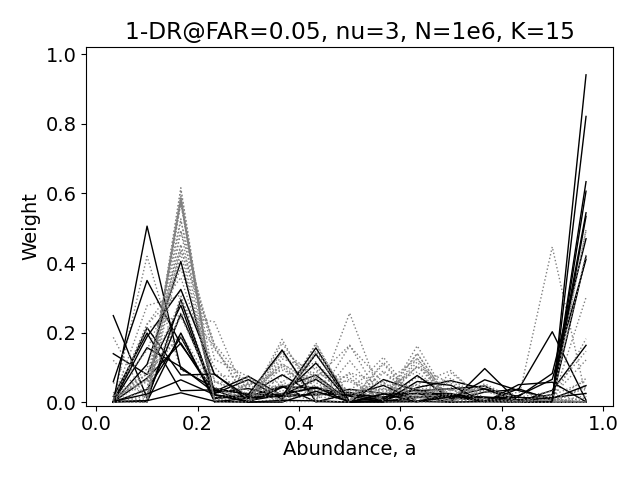}
  \ig[width=0.33\textwidth]{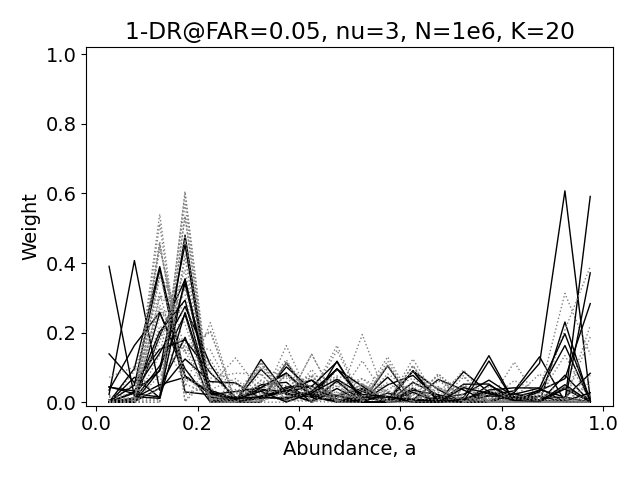}
  \caption{Priors that optimize the DR@FAR=0.05 statistic for increasing number of knots $K$.
    \label{fig:knots-pdx}}
\end{figure}
 \begin{figure}
  \ig[width=0.33\textwidth]{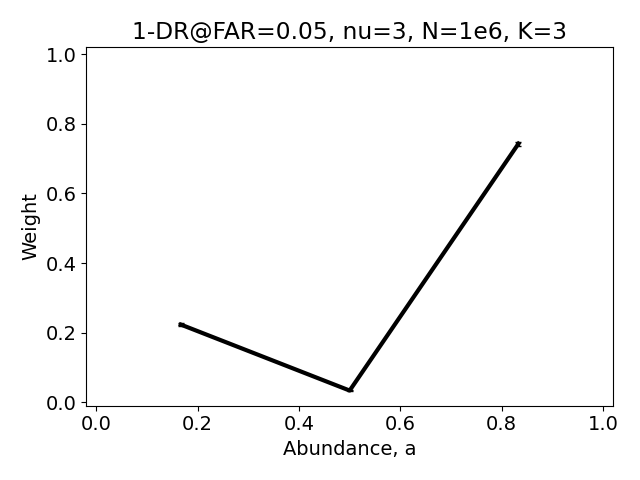}
  \ig[width=0.33\textwidth]{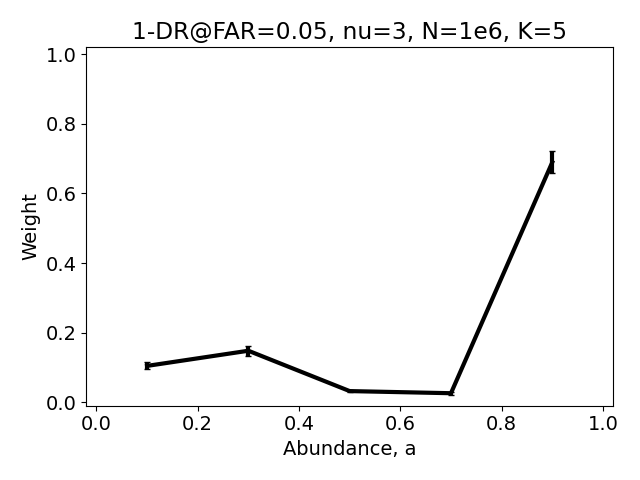}
  \ig[width=0.33\textwidth]{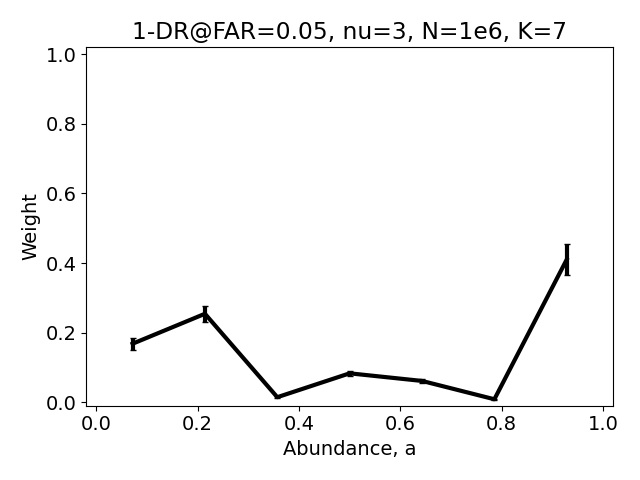} \\
  \ig[width=0.33\textwidth]{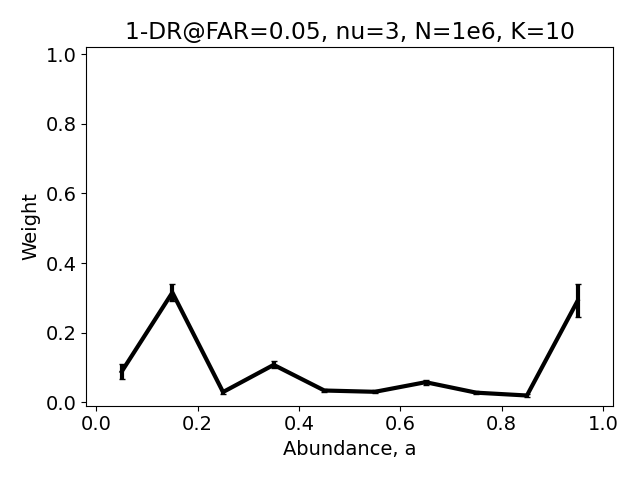} 
  \ig[width=0.33\textwidth]{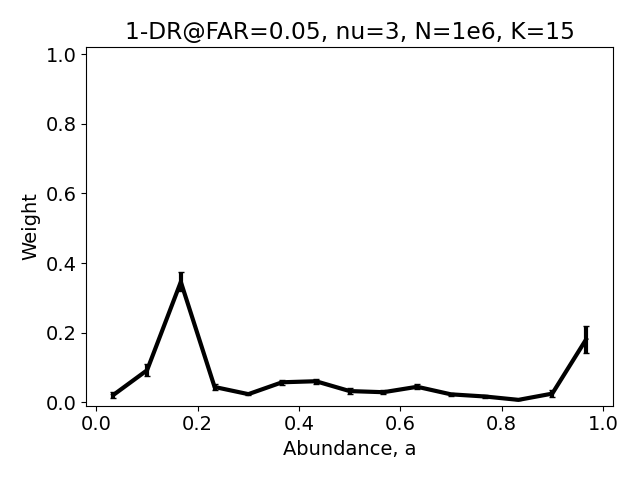}
  \ig[width=0.33\textwidth]{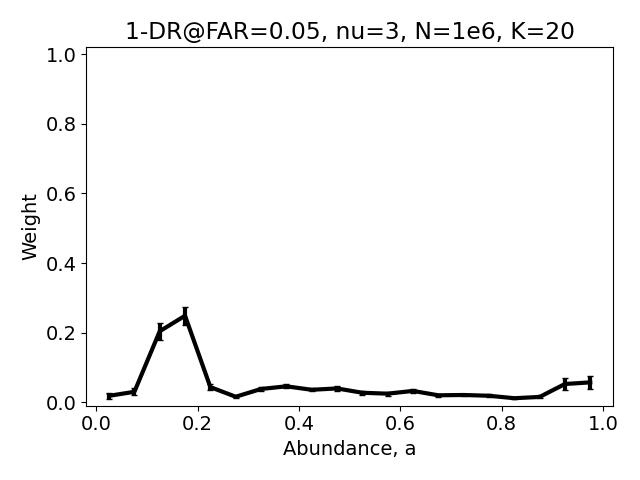}
  \caption{Priors that optimize the DR@FAR=0.05 statistic for increasing number of knots $K$. Shown is a mean of 45 trials, with error bars indicating standard error.
    \label{fig:knots-pdx-mean}}
\end{figure}

 \begin{figure}
  \ig[width=0.33\textwidth]{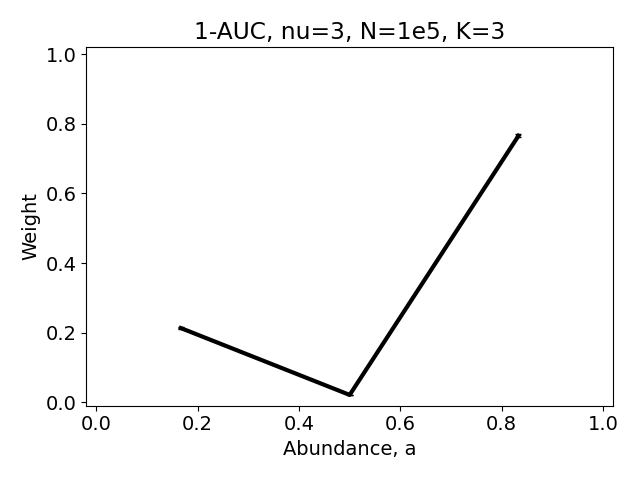}
  \ig[width=0.33\textwidth]{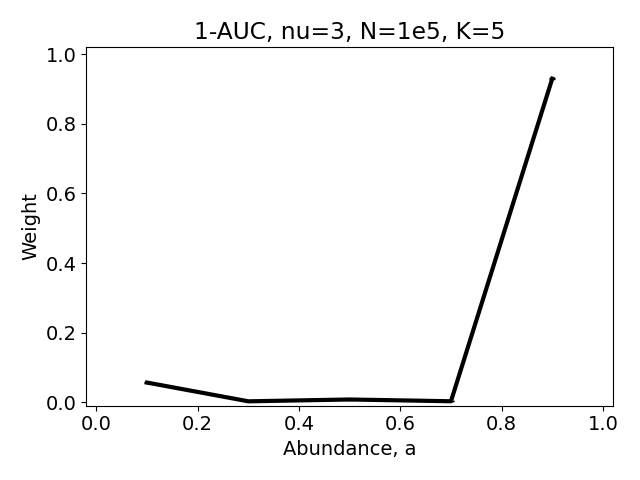}
  \ig[width=0.33\textwidth]{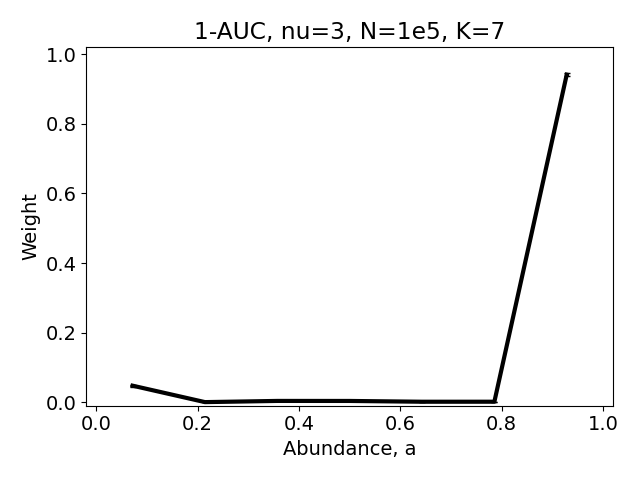}\\
  \ig[width=0.33\textwidth]{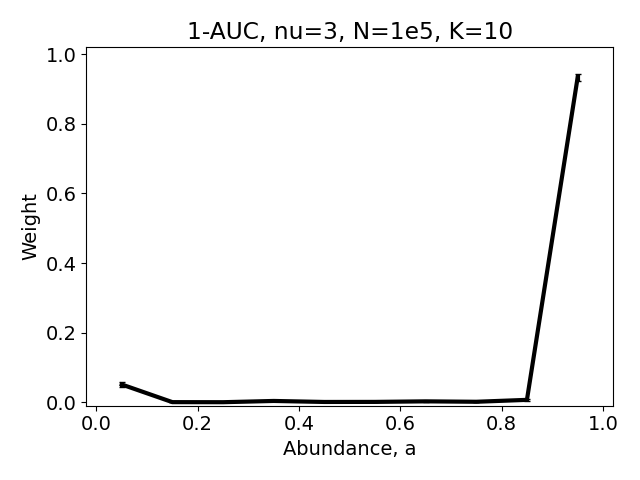}
  \ig[width=0.33\textwidth]{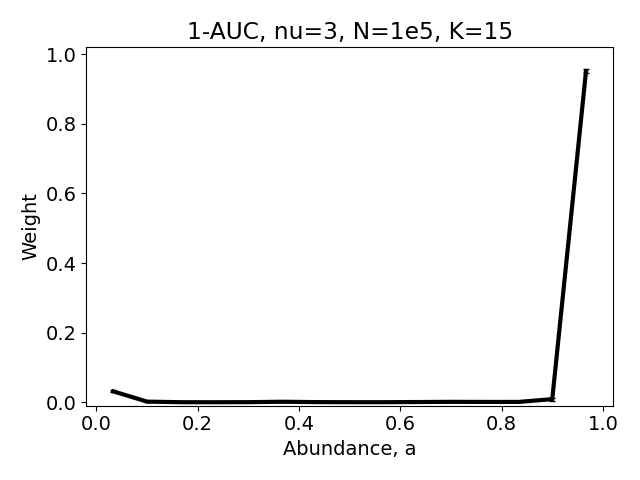}
  \ig[width=0.33\textwidth]{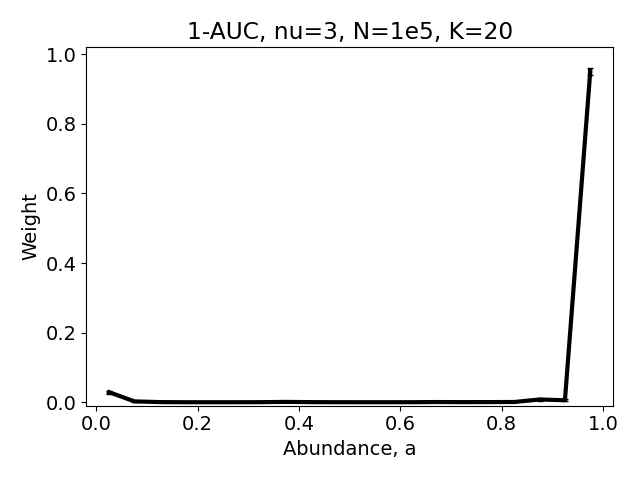}
  \caption{Priors that optimize the AUC statistic for increasing
    number of knots $K$. Shown is a mean of 45 trials, with error
    bars indicating standard error.  These plots suggest that, from the
    point of view of optimizing AUC, a very good detector might be the
    simple likelihood ratio in the limit as $a\to 1$.\remark{Might
      be interesting to see what that looks like in closed form.}
    \label{fig:knots-auc-mean}
}
\end{figure}

\begin{figure}
  \ig[width=0.33\textwidth]{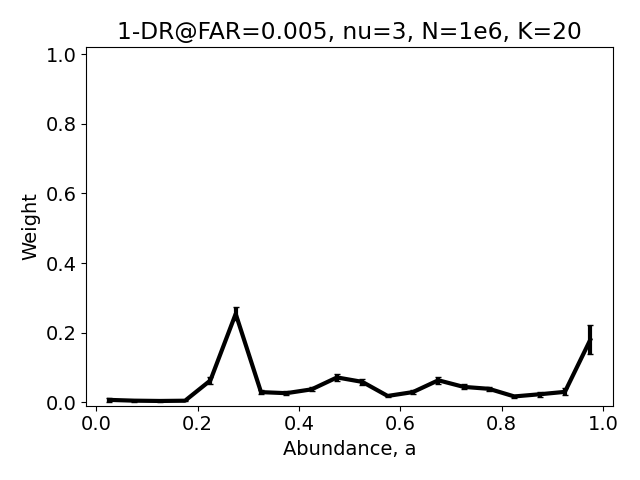}
  \ig[width=0.33\textwidth]{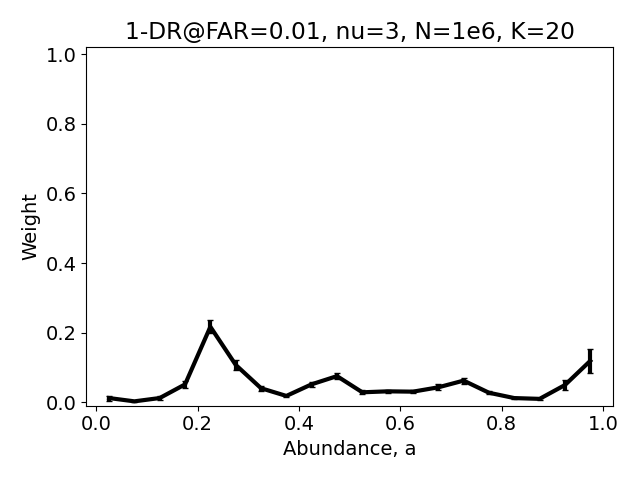}
  \ig[width=0.33\textwidth]{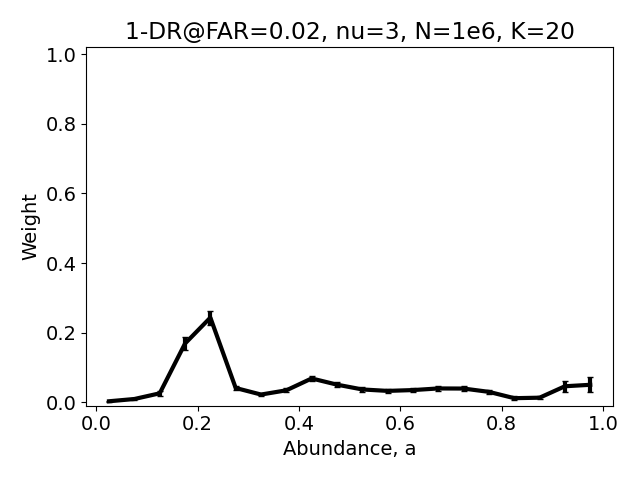}\\
  \ig[width=0.33\textwidth]{pdx-a20-N1e6-D9-wts-mean.png}
  \ig[width=0.33\textwidth]{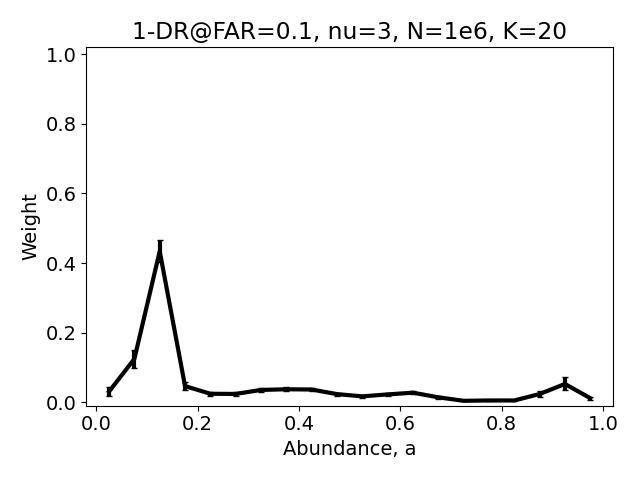}
  \ig[width=0.33\textwidth]{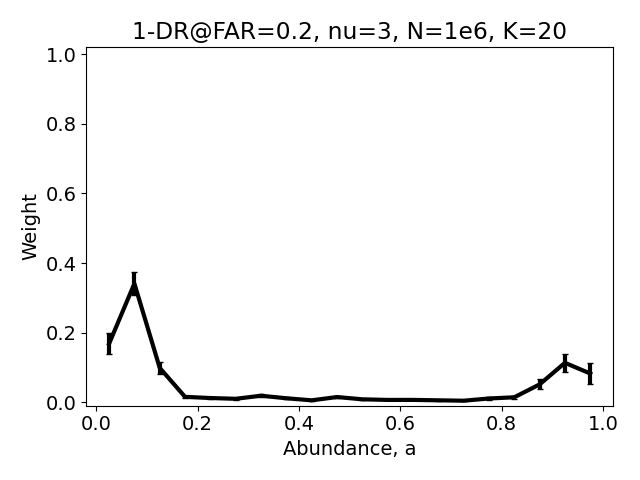}\\
  \ig[width=0.33\textwidth]{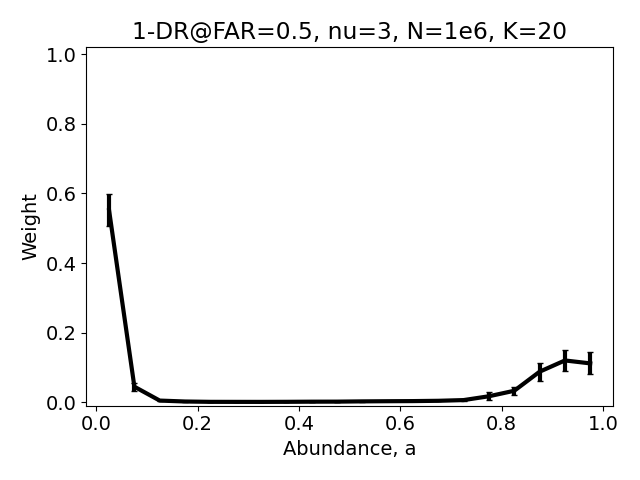}
  \ig[width=0.33\textwidth]{aux-a20-N1e5-D9-wts-mean.png}
  \ig[width=0.33\textwidth]{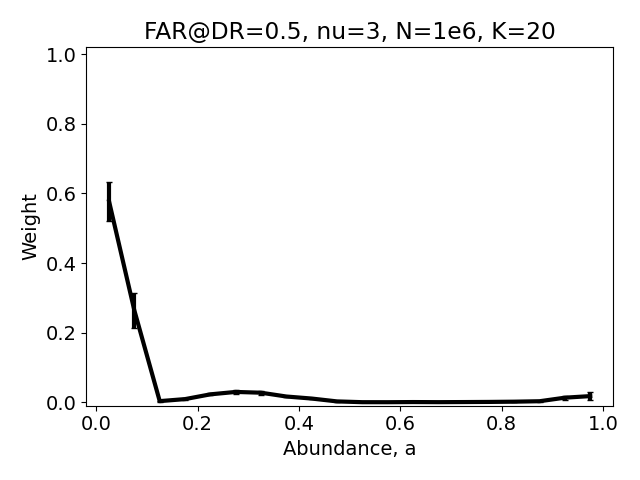}
  \caption{\label{fig:priors-mean-K20}
    Prior weights for various performance criteria; using $K=20$,
    and averaging over 45 trials.}
\end{figure}

The qualitative goal of this study was to see what the priors look
like that enable Bayesian detectors to outperform their non-Bayesian
(GLRT) counterparts.  What was found is that the shape of a
high-performing prior depends on the particular choice of detection
statistic that is being optimized.  In \fig{priors-K5}, nine such
statistics are considered, the first seven of which are of the form DR@FAR=$x$,
for different values of $x$.  Despite the similarity of these
statistics, the shapes of those priors vary considerably for different
values of $x$.  The AUC statistic, seen in the eighth panel of \fig{priors-K5},
and the FAR@DR=0.5 statistic, seen in the ninth panel, exhibits yet other
shapes.  Also evident from these figures:
not only do the shapes vary, but the amount of trial-to-trial
variability in high-performing priors also varies considerably with
choice of statistic.

Although we speak of Bayesian detectors (with appropriately sculpted
priors) being uniformly more powerful than their GLRT counterparts, it
is worth keeping in mind, at least for the experiments reported here,
that the actual performance differences are quite small.
\fig{differences} shows that these detectors exhibit detection rates
that are only one to two percent higher than the optimal clairvoyant
detectors, and that the differences between Bayes and RGLRT is less
than a tenth of a percent.

In assessing the qualitative shape of successful priors, we would
ultimately like to consider continuous functions $q(a)$, but for
numerical reasons we stick to the discrete case with a bounded number
$K$ of knots.  \fig{knots-pdx} shows what the optimal prior looks like
for the DR@FAR=0.05 statistic, and we observe 
increasingly complicated curves for increasing $K$. The
general trend for all these trials is to put most of the weight near
$a=0.18$ or so, with higher weights also near $a=1$, but also some
weight for a range of $a$ between those values. To some extent, the
jaggedness of these curves reflects the numerical challenges of
optimization in a space of high dimension -- that dimension is $K-1$,
arising from the $K$ weights, minus the the one constraint that
weights sum to one. The trend is easier to see in
\fig{knots-pdx-mean}, which is the average of the 45 trials.

A contrasting situation is seen in \fig{knots-auc-mean}, which
considers the same question for the AUC statistic. In this case, the
curves appear to be approaching an almost singular function with
almost all of the weight concentrated near $a=1$.  As noted in
Ref.~[\citenum{Theiler21veritas}] (see Eq.(28) of that reference), we
can express the detector associated with $a\to 1$ in closed form.

Finally, in \fig{priors-mean-K20}, the mean sculpted priors for $K=20$ are
shown for the same nine ROC statistics that were used in \fig{priors-K5}.

\section{Concluding Discussion}

Admissibility is a good thing, but the fact that Bayesian decision
rules are admissible doesn't mean Bayesian detectors are \emph{always}
better; this was shown previously\cite{Theiler23igarss} in the context of
FAR@DR=$x$ statistics. What is shown here is that even when Bayesian
detectors \emph{are} better (as is the case for DR@FAR=$x$ and AUC
statistics), this theoretical superiority comes with some practical
disadvantages.  What we've observed in the experiments shown here is
that when Bayesian detectors are better than the non-Bayesian GLRT or
RGLRT detectors, they are better by a tiny amount.  Furthermore, in
order to achieve these tiny advantages, the priors must be very
carefully (and expensively) sculpted -- and a prior that works (\ie,
that achieves an incrementally higher detection rate than the
associated non-Bayesian detector) for one criterion will not work for
a related criterion.

On the other hand, we can turn this argument around and point out that
the difference between two Bayesian detectors, which differ in their
choice of prior, is often fairly small, so that a detector that is
built from a generic non-sculpted prior may still be useful.  A
particular example is the \emph{veritas} detector, which is Bayesian
with a single delta-function prior.  One practical
advantage of this detector is that the Bayesian integral is
straightforward can can lead to simple closed-form expressions for the
detector, even when the background model is relatively complicated.

A second advantage to Bayesian detectors is that they work well in a
matched-pair machine learning scenario\cite{Theiler13mpml}.  In this
case, we do not have an explicit model for $\Pbkg(\x)$, but instead
use discriminative machine learning to distinguish between $\Pbkg(x)$
and $\Ptgt(a,x)$. In this scenario, we have a population of
target-free data (the original pixels in the image) and a
corresponding population of pixels into which targets have been
implanted. The sculpted prior provides a distribution for target
strengths $a$ to be used in generating the implanted target pixels.

\section{ACKNOWLEDGMENTS}

I am grateful: to the Laboratory Directed Research and Development
(LDRD) program at Los Alamos National Laboratory for support; to
Stefania Matteoli for valuable discussions about Bayesian detection;
and to Tory Carr for asking the question, after seeing a talk I gave
on Ref.~[\citenum{Theiler23igarss}], ``Well, what \emph{would} those priors
look like?''

\appendix
\section{Some technical details}

Because the performance differences (as seen in \fig{differences} for instance)
are so small, it takes very large sample sizes $N$ (upwards of
ten million samples) in order to determine which detector achieved
the best performance. Even with these large sample sizes, considerable
variability was observed from one batch of samples to the next.  In general,
45 batches were used for each experiment so that this variability could be
assessed (and averaged over).

The numerical problem of finding weights $w_1,\ldots,w_K$ to optimize
the given performance statistic was more difficult than originally
expected.  Evaluating the performance for a given set of weights is
somewhat expensive, with $O(NK^2)$ steps required.  (One factor of $K$
arises from the $K$ terms in the Bayesian sum, and the other factor
from the need to evaluate that sum for each of the $K$ knots.)
Further, the evaluation only provides a scalar performance metric, so
optimization schemes that require a gradient cannot be directly
employed. In terms of the optimizers in the {\tt scipy.optimize}
tools\footnote{\url{https://docs.scipy.org/doc/scipy/reference/optimize.html}},
this reduces to: Powell's method, Nelder-Mead (sometimes called the
``simplex'' method), and COBYLA (Constrained Optimization BY Linear
Approximation).  For the sculpting priors problem, all three of these
exhibited inconsistent convergence (particularly for larger values of
$K$), though COBYLA was much faster, so some initial results were
obtained by using many random initial guesses, applying COBYLA to
each, and choosing the one with the smallest optimum value.

For the results shown here, however, a klunky home-grown optimizer
(called ``babysteps'') was employed. \remark{Should probably point to
  the code, which should probably be incorporated into bluehat}
Although we do not have a
gradient, \textit{per se}, we do know that increasing the weight $w_k$
at a given knot $a_k$ will improve the performance of the
Bayesian detector at that knot. The idea is to increase the weight
associated with the knot that most needs it.  This increase is by a
small amount (a baby step), and all the weights are subsequently
rescaled so the sum of weights is equal to one.  This process of getting
a new set of weights is then iterated.  In this work a fixed
step size and fixed number of iterations were used.

Using a specified ROC-based statistic, the RGLRT detector is evaluated for
each target strength $a_1,\ldots,a_K$. Since RGLRT does not depend on weights,
this evaluation, which provides $s_{\mbox{\scriptsize RGLRT}}(a_k)$ for each $a_k$
need only be done once.

For each iteration of the babysteps algorithm, we begin with a set of
weights $\w=[w_1,\ldots,w_K]$, and evaluate the Bayesian detector at
each $a_1,\ldots,a_k$.  Since our loss function is of the form $L(\w)
= \max_k L(\w,a_k)$ as seen in \eq{loss},
it is clear that $k_* = \mbox{argmax}_k L(w,a_k)$ is the knot
most in need of improvement.  So we update the weights with
$w_{k_*} \leftarrow
w_{k_*} + \Delta w$ (where typically $\Delta w = 0.01$), followed by
$w_k \leftarrow w_k /(1+\Delta w)$ for all $k=1,\ldots,K$.

\begin{figure}
  \ig[width=0.33\textwidth]{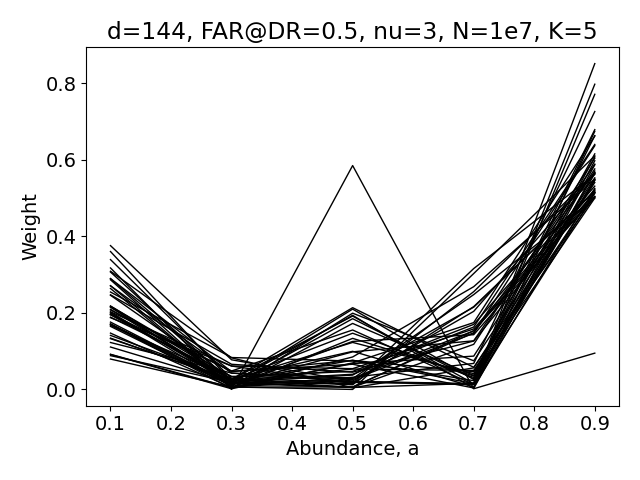}
  \parbox[b]{0.66\textwidth}{\caption{Example in which priors can be
      found that lead to uniformly more powerful Bayesian detectors
      using the FAR@DR=0.5 statistic.  This is similar to the last
      panel in \fig{priors-K5} except that $d=144$ instead of
      $d=9$. Here, all the curves are solid (Bayes wins), while in
      \fig{priors-K5}, the curves are dotted (indicating that Bayes is
      \emph{not} uniformly better).\\ {~}\\{~}\\
    \label{fig:priors-FAR-D144}}}
\end{figure}

As was briefly mentioned in the caption to \fig{priors-K5}, with
respect to the theory guaranteeing the existence of priors that lead
to uniformly superior behavior for Bayesian detectors, that theory
does not apply for the FAR@DR=0.5 detector (the reason for that is
explained in Ref.~[\citenum{Theiler23igarss}]).  The theory, however, does
not rule out the possibility that those priors might exists.  And
although they do not appear to exist for the case shown in the last
panel of \fig{priors-K5}, they have been seen in other cases.  One such
example is shown in \fig{priors-FAR-D144}, which differs from the
examples in \fig{priors-K5} in the choice of the spectral dimension:
$d=9$ in \fig{priors-K5}, and $d=144$ in \fig{priors-FAR-D144}. (Note
that there is no extra computation time or memory expense in using the
larger spectral dimension because all of the detectors can be expressed in
two-dimensional matched-filter-residual (MFR) coordinates\cite{Foy09}.)

\bibliography{sculpt}
\bibliographystyle{spiebib}

\end{document}